\documentclass[twocolumn,showpacs]{revtex4}
\usepackage{eurosym}
\usepackage{graphicx}
\usepackage{dcolumn}
\usepackage{bm}
\usepackage{amsmath}
\usepackage{amsfonts}
\usepackage{amssymb}
\providecommand{\U}[1]{\protect\rule{.1in}{.1in}}

\begin{document}
\title{The Fibonacci quantum walk and its classical trace map}
\author{A. Romanelli}
\altaffiliation{\textit{E-mail address:} alejo@fing.edu.uy}
\affiliation{Instituto de F\'{\i}sica, Facultad de Ingenier\'{\i}a\\
Universidad de la Rep\'ublica\\
casilla de correo 30, c\'odigo postal 11000, Montevideo, Uruguay}
\date{\today }

\begin{abstract}
We study the quantum walk in momentum space using a coin arranged in
quasi-periodic sequences following a Fibonacci prescription. We
build for this system a classical map based on the trace of the
evolution operator. The sub-ballistic behavior of this quantum walk
is connected with the power-law decay of the time correlations of
the trace map.
\end{abstract}

\maketitle

\section{Introduction}
The quantum walk (QW) on the line is a subject that has drawn much
attention in quantum computing \cite{Kempe}. It has been introduced by \cite%
{Aharonov,Godoy}, as counterpart of the classical random walk. Due
to the success of the classical random walks in the description of
many classic models of the nature, the possibility that future
quantum algorithms will be based on the QW has attracted the
attention of researchers from different fields. Additionally, as a
result of recent advances in experimental techniques to manipulate
quantum states has led the scientific community to think that
quantum computers could be a reality in the near future. Anyway the
study of the QW subjected to different sources of decoherence is a
topic that allows to analyze and verify the principles of quantum
theory and has been considered by several authors \cite{kendon}.

One of the most striking properties of the QW is its ability to
spread over the line linearly in time as characterized by the
standard deviation $\sigma (t)\sim t$, while its classical analog
spreads out as the square root of time [$\sigma (t)\sim t^{1/2}$].
Recently, in ref. \cite{Ribeiro} the QW was generalized introducing
two coin operators arranged in quasi-periodic sequences following a
Fibonacci prescription. This \textquotedblleft Fibonacci QW" leads
to an unexpected sub-ballistic wave function spreading, as shown by
the power-law tail of the standard deviation [$\sigma (t)\sim t^{c}$
with $0.5<c<1$]. More recently\cite{alejo0,alejo4}, we have studied
the QW subjected to noise with a L\'{e}vy waiting-time distribution
\cite{Levy}, finding that the system has also a sub-ballistic wave
function spreading instead of the known ballistic growth. The
Fibonacci sequence gives rise to a rich behavior also in other
systems; it is the cause of sub-ballistic behavior both for the
quantum kicked rotor in resonance \cite{alejo3} and for
tight-binding electrons \cite{Abe,Zhong}. On the other hand
Sutherland \cite{Sutherland}, twenty years ago, studied a very
simple two dimensional system of a single particle with spin $1/2$
subjected to quasi-periodic external field pulses, showing that the
dynamics of the system is determined by a three dimensional
\textquotedblleft trace map". In that work, he found three regions
of the parameters space characterized by qualitative differences of
the decay of the time-correlation function: slower than a power law,
as a power law, and faster than a power law.

Here we study the Fibonacci QW through the dynamics of its trace
map. We show that the sub-ballistic behavior of the QW is connected
with the power- law decay of the time-correlation function of the
trace map.

\section{Fibonacci quantum walk}

\label{walked}The dynamics of the Fibonacci QW will be generated by
a large sequence of two time-step unitary operators $U_{1}$ and
$U_{2}$ for each time $t$. Given the two initial values of the
succession $U_{1}$ and $U_{2}$
the sequence is obtained, as was done in previous works \cite%
{Kohmoto,Sutherland,Ribeiro}, using the rule
\begin{equation}
U_{k+1}=U_{k}U_{k-1}.  \label{evol}
\end{equation}
To obtain the operators $U_{1}$ and $U_{2}$ we develop in some
detail the free QW model. The standard QW corresponds to a
one-dimensional evolution of a quantum system (the walker) in a
direction which depends on an additional
degree of freedom, the chirality, with two possible states: ``left" $%
|L\rangle $ or ``right" $|R\rangle $. Let us consider that the walker can
move freely over a series of interconnected sites labeled by an index $n$.
In the classical random walk, a coin flip randomly selects the direction of
the motion; in the QW the direction of the motion is selected by the
chirality. At each time step a rotation (or, more generally, a unitary
transformation) of the chirality takes place and the walker moves according
to its final chirality state. The global Hilbert space of the system is the
tensor product $H_{s}\otimes H_{c}$ where $H_{s}$ is the Hilbert space
associated to the motion on the line and $H_{c}$ is the chirality Hilbert
space.

We shall consider only unitary transformations which can be
expressed in terms of a single real angular parameter $\theta $
\cite{Nayak,Tregenna,Bach}. Let us call $T_{-}$ ($T_{+}$) the
operators that move the walker one site to the left (right) on the
line in $H_{s}$ and let $|L\rangle \langle L|$ and $|R\rangle
\langle R|$ be the chirality projector operators in $H_{c}$. Then we
consider free evolution transformations of the form \cite{Nayak},
\begin{equation}
U_{i}(\theta _{i})=\left\{ T_{-}\otimes |L\rangle \langle L|+T_{+}\otimes
|R\rangle \langle R|\right\} \circ \left\{ I\otimes K(\theta _{i})\right\} ,
\label{Ugen}
\end{equation}%
where $K(\theta _{i})=\sigma _{z}e^{-i\theta _{i}\sigma _{y}}$ is an
unitary operator acting on $H_{c}$, $\sigma _{y}$ and $\sigma _{z}$
being the standard Pauli matrices, and $I$ is the identity operator
in $H_{s}$. The unitary operator $U_{i}(\theta _{i})$ evolves the
state $|\Psi (t)\rangle $ by one time step,
\begin{equation}
|\Psi (t+1)\rangle =U_{i}(\theta _{i})|\Psi (t)\rangle .  \label{evol1}
\end{equation}%
The wave vector $|\Psi (t)\rangle $ is expressed as the spinor
\begin{equation}
|\Psi (t)\rangle =\sum\limits_{n=-\infty }^{\infty }{\binom{a_{n}(t)}{%
b_{n}(t)}}|n\rangle ,  \label{spinor}
\end{equation}%
where we have associated the upper (lower) component to the left (right)
chirality, the states $|n\rangle $ are eigenstates of the position operator
corresponding to the site $n$ on the line. The unitary evolution for $|\Psi
(t)\rangle $, corresponding to Eq.~(\ref{evol1}) can then be written as the
map
\begin{align}
a_{n}(t+1)& =a_{n+1}(t)\,\cos ~\theta +b_{n+1}(t)\,\sin ~\theta \,,
\label{mapa} \\
b_{n}(t+1)& =a_{n-1}(t)\,\sin ~\theta -b_{n-1}(t)\,\cos ~\theta \,.  \notag
\end{align}%
To build the operators $U_{1}$ and $U_{2}$ we substitute in the
previous expression $\theta $ by $\theta _{1}$ and $\theta _{2}$
respectively. As mentioned in the introduction these types of maps
produce a sub-ballistic wave function spreading.

\section{Trace map}

\label{traced} With the operators defined in the previous section,
we shall study the Fibonacci QW connecting it explicitly with the work of Sutherland \cite%
{Sutherland}. To this end, we make the spatial Fourier transform of
the amplitude $(a_{n}(t),b_{n}(t))^{T}$ multiplying both sides of
Eq.~(\ref{mapa}) by $e^{i(\phi-\pi/2) n}$, with $\phi \in \left[
-\pi ,\pi \right] $, and summing in the integer index $n$:
\begin{equation}
{\binom{F(\phi ,t+1)}{G(\phi ,t+1)}}=M(\phi ,\theta ){\binom{F(\phi ,t)}{%
G(\phi ,t)\,}},  \label{mapa2}
\end{equation}%
where
\begin{align}
F(\phi ,t)& =\sum\limits_{n}{e^{i n(\phi-\pi/2) }}a_{n}(t)\,,  \label{efe} \\
G(\phi ,t)& =\sum\limits_{n}e^{i n(\phi-\pi/2) }b_{n}(t)\,,
\label{ge}
\end{align}%
and
\begin{equation}
M(\phi ,\theta )=i\left(
\begin{array}{cc}
e^{-i\phi }\cos ~\theta  & e^{-i\phi }\sin ~\theta  \\
e^{i\phi }\sin ~\theta  & -e^{i\phi }\cos ~\theta
\end{array}%
\right) {.}  \label{matriz}
\end{equation}%
Thus in Fourier space, the dynamics of the Fibonacci QW is
determined by the $2\times 2$ unitary matrix $%
M(\phi ,\theta )$. We call $M_{1}$ and $M_{2}$ the matrix $M(\phi
,\theta )$ evaluated in $\left( \phi _{1},\theta _{1}\right) $ and
$\left( \phi _{2},\theta _{2}\right) $ respectively. The matrix
$M(\phi ,\theta )$ can be rewritten employing the
Bloch-sphere representation as%
\begin{equation}
M(\phi ,\theta )=\cos \theta \sin \phi \text{ }I+i\overset{%
\rightarrow }{u}.\overset{\rightarrow }{\sigma } , \label{matriz2}
\end{equation}%
where $I$ is the identity matrix, $\overset{\rightarrow }{u}$ is
given by
\begin{equation}
\overset{\rightarrow }{u}{=}\left( \sin ~\theta \cos \phi ,\sin \theta \sin
\phi ,\cos \theta \cos \phi \right) ,  \label{versor}
\end{equation}%
and $\overset{\rightarrow }{\sigma }=(\sigma _{x},\sigma _{y},\sigma
_{z})$. The vector $\overset{\rightarrow }{u}$ can point in any
direction but for our purposes only the relative orientation
between $\overset{\rightarrow }{u}_{1}$ and $\overset{%
\rightarrow }{u}_{2}$ is important. If we choose $\phi _{1}=\pi /2$
then
\begin{equation}
M_{1}=\cos \theta_{1} \text{ }I+i\sin \theta_{1} \text{ }\sigma
_{y}.
\end{equation}%
The values of the parameters $\theta _{1}$, $\theta _{2}$ and $\phi
_{2}$ determine the initial state for the dynamics of the system. In
the Fourier space, the system can be
interpreted as a particle with spin $1/2$ subjected to two types of pulses, $%
M_{1}$ and $M_{2}$, coming from two external magnetic fields \cite%
{Sutherland}. The dynamics of the spin is specified by the order of
the interaction pulses. In our case the pulse $M_{1}$ is applied
first, then the pulse $M_{2}$ is applied and the following pulses
are determined by the rule
\begin{equation}
M_{k+1}=M_{k}M_{k-1}.  \label{evol2}
\end{equation}%
The pulse $M_{1}$ is a rotation with axis pointing in the positive
$\overset{\rightarrow }{j}$ direction, then it is completely
determined specifying
\begin{equation}
x_{1}=Tr M_{1}/2=\cos \theta _{1},  \label{y}
\end{equation}
where $Tr A$ is the trace of the matrix $A$. The pulse $M_{2}$ can
point in any direction but it is determined specifying $\theta _{2}$
and $\phi _{2}$; or equivalently by
\begin{eqnarray}
y_{1} &=&Tr M_{2}/2=\sin \phi _{2}\cos \theta _{2},  \label{x} \\
z_{1} &=&Tr(M_{2}M_{1})/2=\sin \phi _{2}\cos (\theta _{2}+\theta
_{1}). \label{z}
\end{eqnarray}%
\begin{figure}[h]
\begin{center}
\includegraphics[scale=0.42]{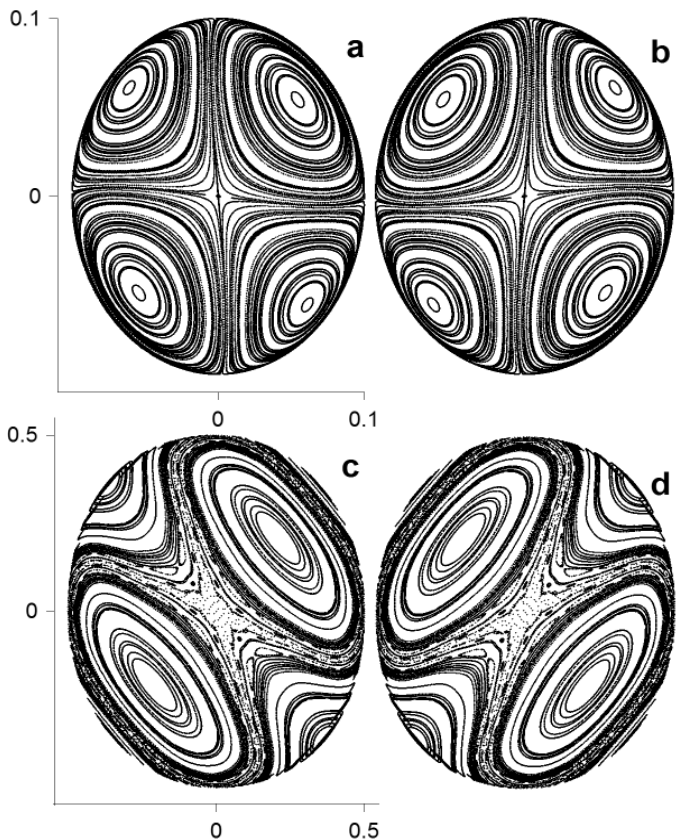}
\end{center}
\caption{Poincare sections for the two hemispheres of the trace map.
Arbitrary units are used. For C=-0.99  a) back and b) front
hemispheres. For C=-0.7 c) back and d) front hemispheres.}
\label{fig1}
\end{figure}
\begin{figure}[h]
\begin{center}
\includegraphics[scale=0.42]{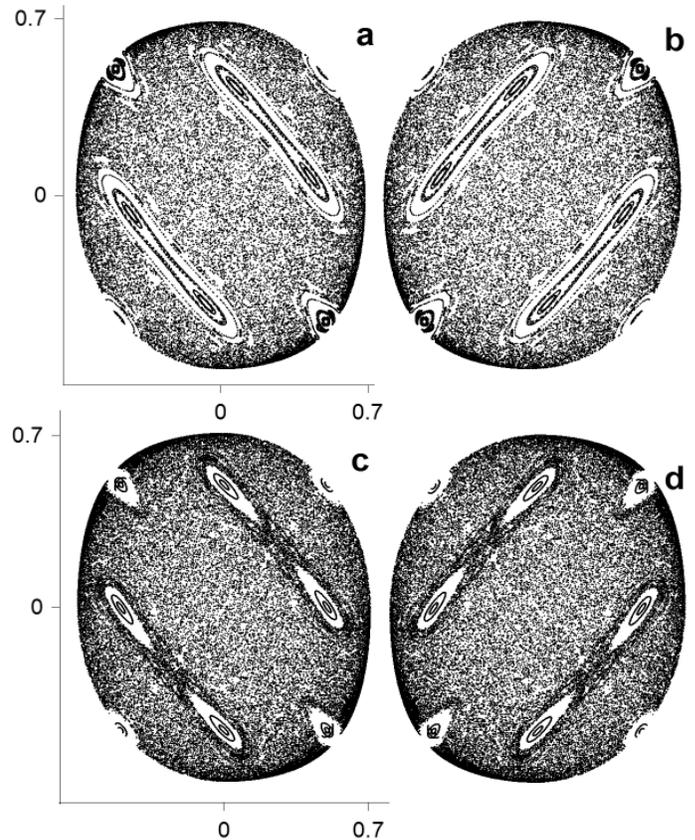}
\end{center}
\caption{Poincare sections for the two hemispheres of the trace map.
Arbitrary units are used. For C=-0.53  a) back and b) front
hemispheres. For C=-0.5 c) back and d) front hemispheres.}
\label{fig2}
\end{figure}
Starting from the ordered pair \text{ }$\left( M_{1},M_{2}\right) $,
called standard alignment, any other orientation $\left(
M_{k},M_{k+1}\right) $ can be thought as a rotation of $SU(2)$. That
is
\begin{equation}
\left( M_{k},M_{k+1}\right) =R\left( M_{1},M_{2}\right) R^{-1},
\label{rotation}
\end{equation}%
where $R$ is the rotation operator. The vector $%
\overset{\rightarrow }{r_{k}}=\left( x_{k},\text{ }y_{k},\text{ }%
z_{k}\right) $ is defined as
\begin{equation}
\overset{\rightarrow }{r_{k}}\equiv \frac{1}{2}\left( Tr M_{k},\text{ }%
Tr M_{k+1},\text{ }Tr M_{k+2}\right) .  \label{r}
\end{equation}%
Now, the dynamics of the system can be visualized through the \
iteration of the trace map $\left( x_{k},\text{ }y_{k},\text{
}z_{k}\right) \Longrightarrow $ $\left( x_{k+1},\text{
}y_{k+1},\text{ }z_{k+1}\right) $ with initial condition $\left(
x_{1},\text{ }y_{1},\text{ }z_{1}\right) $. Using some properties of
these matrices, Kohmoto \textit{et al.} \cite{Kohmoto} showed that
the vector $\overset{\rightarrow }{r_{k}}$ satisfies the following
nonlinear map
\begin{equation}
\left( x_{k+1},\text{ }y_{k+1},\text{ }z_{k+1}\right) =\left( y_{k},\text{ }%
z_{k},\text{ }2y_{k}z_{k}-x_{k}\right) .  \label{mapar}
\end{equation}%
This map has an invariant given by
\begin{equation}
C\equiv x_{k}^{2}+y_{k}^{2}+z_{k}^{2}-2x_{k}y_{k}z_{k}-1=-\left( \sin \theta
_{1}\cos \phi _{2}\right) ^{2}.  \label{invariante}
\end{equation}%
The value of the invariant is obtained by substitution of the initial
condition in the generic expression of $C$. It also follows from Eq.~(\ref%
{invariante}) that $-1\leq C\leq 0$. In Fig.~\ref{fig1} and
Fig.~\ref{fig2} we show a sequence of Poincare sections of the trace
map in the plane $(x,z)$ for increasing values of $C$. This sequence
shows that the size of the sections grows with the value of the
invariant $C$. As the invariant surface has the topology of a sphere
we represent separately both hemispheres corresponding to positive
and negative values of $y$. The typical orbits of the map are
presented over both hemispheres. For each value of $C$ the left
(right) figure is the back (front) hemisphere. Both hemispheres look
similar, as if there existed a symmetry operation between them. Each
hemisphere presents several axes of symmetry and some elliptic and
hyperbolic points.

In Fig.~\ref{fig1},a,b for $C=-0.99$ there appear to be no chaotic
regions for the scale shown. It is easy to identify four elliptic
points and one hyperbolic point with different winding numbers.
Therefore, for values of $C$ near $-1$ there appears to be a second
integral of the motion, at least to the accuracy of these plots, and
the system seems to be integrable.

In Fig.~\ref{fig1},c,d for $C=-0.7$, the largest resonances are
clearly seen, but some of the periodic orbits are pushed to the
periphery. The neighborhood of the central hyperbolic fixed point
begins to be chaotic but the plot is still dominated by the KAM
tori.

In Fig.~\ref{fig2},a,b for $C=-0.53$, the central hyperbolic fixed
point has a chaotic sea but there are still some elliptic points
stretching diagonally.

Fig.~\ref{fig2},c,d for $C=-0.5$, is similar to the previous case
but now the each elliptic zone breaks into two pieces and the
chaotic sea increases.

\begin{figure}[h]
\begin{center}
\includegraphics[scale=0.30]{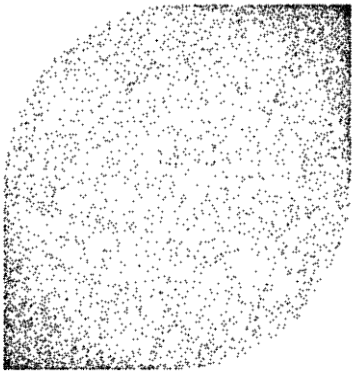}
\end{center}
\caption{Front hemisphere for the trace map whit C=0, the back
hemisphere looks the same up to a rotation.} \label{fig3}
\end{figure}

It is interesting to study the trace map when $\phi_2=\pi/2$, and
then $C=0$. As showed in Fig.~\ref{fig3} the second integral of
motion appears to have been totally destroyed and the integrability
of the system is lost. In this case the invariant surface has two
accumulation points. In the other extreme, when $C=-1$, the solution
of the trace map is only one point.

For intermediate values of $C$ chaos sets in and large regions
contain a chaotic sea of trajectories. There is no way to predict
the future evolution of a trajectory in these regions, but it is
possible to use the techniques of stochastic theory to determine the
statistical behavior of trajectories. If no stable islands or
cantori exist, the diffusion process is like that found in simple
Brownian motion. However, if stable islands and cantori exist, the
diffusion process becomes much complex. Here we see that there are
many stable islands embedded in the chaotic region. The darker
shading of trajectories along some lines indicate that these regions
are bracketed by cantori. During the wandering in the chaotic sea
the trajectory could get trapped in the neighborhood of a stable
island for arbitrary lengths of time. The trapping of trajectories
by the stable islands gives rise to power-law decay of trajectory
correlations rather than exponential decays as would be expected for
a purely random diffusion process. Sutherland \cite{Sutherland} has
computed the correlation function for the trace map.
Our equations are the same as in Ref. \cite{Sutherland}, but our
initial conditions are chosen obeying the initial conditions of the
Fibonacci QW. In both models: i)the identification of the most
obvious elliptic and hyperbolic $K$-cycles are similar, ii) he
relation between the rotation of the standard alignment $(M_1,M_1)$
and the $K$-cycles of the trace map are common, iii) the trace map
dynamics determines the rotation dynamics. Therefore we can
translate the results of the Sutherland model to the Fourier
transform of the Fibonacci QW. Then, here also three regions of the
parameter space are characterized by the qualitative differences in
the decay of the time correlations. These regions are: i) near a
stable elliptic cycle where the decay is slower than a power of
time. ii) near a hyperbolic cycle where the decay is a power of
time. iii) in the chaotic region where the decay is faster than a
power of time.

\section{Conclusion}

\label{sec:conclusion} In this work, we have established a trace map
for the Fibonacci QW. This map has an integral of the motion $C$
that determines a three dimensional surface where the trajectories
lie. The trajectories depend on the parameter $C$, when this
parameter goes to $-1$ the system is integrable, but when it goes to
$0$ the trajectories become chaotic. For intermediate values of $C$
the trajectories have a rich behavior where it is possible to see on
the $(x,z)$ plane elliptic orbits, hyperbolic orbits, KAM tori,
cantorus, etc. It is in this zone where the autocorrelation function
has a time power-law dependence. This deviation from exponential
decay appears because the stable islands in the chaotic sea can trap
trajectories for long intervals of time. This stickiness of the
islands seems to be due to a network of cantori and higher order
stable islands chains which surround the stable islands and extend
into the chaotic sea. It is possible to show, as in Ref.
\cite{alejo5}, that the sub-ballistic increase in the variance of
the Fibonacci QW is a direct consequence of the coherence of the
quantum evolution. Then, not all the trajectories of the trace map
contribute with the same weight in the Fourier antitransformation of
Eqs.~(\ref{efe},\ref{ge}). The more regular behavior has more weight
than the chaotic trajectories and this leads to the power-law decay.
Therefore, the power-law behavior in the trace map affects the wave
function in such a way to obtain the sub-ballistic behavior of the
standard deviation. However, it is clear that more work will be
necessary to understand these mechanism more deeply.

\section{Acknowledgments}

\label{sec:Acknowledgments} I thank V. Micenmacher  for his comments
and stimulating discussions. I acknowledge support from PEDECIBA and
PDT S/C/IF/54/5.

\end{document}